\documentclass{article}

\usepackage[preprint]{nips_2018}

\usepackage[utf8]{inputenc} 
\usepackage[T1]{fontenc}    
\usepackage{hyperref}       
\usepackage{url}            
\usepackage{booktabs}       
\usepackage{amsfonts}       
\usepackage{nicefrac}       
\usepackage{microtype}      
\usepackage{amsmath}
\usepackage{amssymb}
\usepackage{graphicx}
\usepackage{subcaption}
\usepackage{amsthm}
\usepackage[english]{babel}
\usepackage[toc,page]{appendix}
\usepackage{thm-restate}
\usepackage{cleveref}

\usepackage[colorinlistoftodos,bordercolor=orange,backgroundcolor=orange!20,linecolor=orange,textsize=scriptsize]{todonotes}

\title{Deep learning in bioinformatics: introduction, application, and perspective in big data era}

\author{
  Yu Li \\
  KAUST \\
  CBRC \\
  CEMSE \\
  \And
  Chao Huang \\
  NICT \\
  CAS \\
  \And
  Lizhong Ding \\
  IIAI \\
  \And
  Zhongxiao Li \\
  KAUST \\
  CBRC \\
  CEMSE \\
  \AND
  Yijie Pan \\
  NICT \\
  CAS \\
  \And
  Xin Gao \thanks{All correspondence should be addressed to Xin Gao (xin.gao@kaust.edu.sa).} \\
  KAUST \\
  CBRC \\
  CEMSE \\
}

\begin{document}

\maketitle

\begin{abstract}
Deep learning, which is especially formidable in handling big data, has achieved great success in various fields, including bioinformatics. With the advances of the big data era in biology, it is foreseeable that deep learning will become increasingly important in the field and will be incorporated in vast majorities of analysis pipelines. In this review, we provide both the exoteric introduction of deep learning, and concrete examples and implementations of its representative applications in bioinformatics. We start from the recent achievements of deep learning in the bioinformatics field, pointing out the problems which are suitable to use deep learning. After that, we introduce deep learning in an easy-to-understand fashion, from shallow neural networks to legendary convolutional neural networks, legendary recurrent neural networks, graph neural networks, generative adversarial networks, variational autoencoder, and the most recent state-of-the-art architectures. After that, we provide eight examples, covering five bioinformatics research directions and all the four kinds of data type, with the implementation written in Tensorflow and Keras. Finally, we discuss the common issues, such as overfitting and interpretability, that users will encounter when adopting deep learning methods and provide corresponding suggestions. The implementations are freely available at \url{https://github.com/lykaust15/Deep_learning_examples}.
\end{abstract}

\section{Introduction}
\label{sec:intro}

\begin{table*}[!t]
\textcolor{black}
{\caption{\textcolor{black}{Abbreviations.}}
\centering
\label{tab:abbr}
\begin{tabular}{ p{2cm}p{6cm}p{2cm}}
 \hline
 Abbreviations &  Full words & Reference \\
 \hline
 AUC & Area under curve & \cite{wang2015auc} \\
 \hline
 CNN  & Convolutional neural networks & \cite{RN414} \\
 \hline
 Cryo-EM & Cryogenic electron microscopy & \cite{merk2016breaking} \\
 \hline
 Cryo-ET & Cryogenic electron tomography & \citep{han2018autom} \\
 \hline
 CT & Computed tomography & \cite{kumar2015lung} \\
 \hline
 DenseNet & Densely connected convolutional networks & \cite{RN691} \\
 \hline
 DNN &  Deep fully connected neural networks & \cite{RN630}\\
 \hline
 DPN & Dual path networks & \cite{RN770} \\
 \hline
 DR & Dimensionality reduction & \cite{roweis2000nonlinear}\\
 \hline
 EEG & Electroencephalogram & \cite{RN675} \\
 \hline
 GAN & Generative adversarial networks & \cite{goodfellow2014generative} \\
 \hline
 GCN & Graph convolutional neural networks & \cite{kipf2016semi} \\
 \hline
 RNN  & Recurrent neural networks & \cite{mikolov2010recurrent} \\
 \hline
 ResNet & Residual networks & \cite{RN696} \\
 \hline
 LSTM & Long short-term memory & \cite{graves2013hybrid} \\
 \hline
 MRI & Magnetic resonance imaging & \cite{sarraf2016classification} \\
 \hline
 PET & Positron emission tomography & \cite{li2014deep}\\
 \hline
 PseAAC & Pseudo amino acid composition & \cite{RN493} \\
 \hline
 PSSM & Position specific scoring matrix & \cite{RN775} \\
 \hline
 ReLU & Rectified linear unit & \cite{RN503} \\
 \hline
 SENet & Squeeze-and-excitation networks & \cite{RN728} \\
 \hline
 SGD & Stochastic gradient descent & \cite{zhang2015deep} \\
 \hline
 VAE & Variational auto-encoder & \cite{doersch2016tutorial} \\
 \hline
\end{tabular}
}
\end{table*}

With the significant improvement of computational power and the advancement of big data, deep learning has become one of the most successful machine learning algorithms in recent years \cite{RN630}. It has been continuously refreshing the state-of-the-art performance of many machine learning tasks \cite{RN414,RN676,RN691,RN568} and facilitating the development of numerous disciplines \cite{RN572,RN675,goh2017deep,li2018accelerating}. For example, in the computer vision field, methods based on convolutional neural networks have already dominated its three major directions, including image recognition \cite{RN414,RN696,RN691}, object detection \cite{ren2017faster,redmon2017yolo9000}, and image inpainting \cite{RN532, yu2018generative} and super-resolution \cite{RN679,RN666}. In the natural language processing field, methods based on recurrent neural networks usually represent the state-of-the-art performance in a broad range of tasks, from text classification \cite{zhang2015character}, to speech recognition \cite{RN537} and machine translation \cite{RN568}. Researchers in the high energy physics field have also been using deep learning to promote the search of exotic particles \cite{RN572}. In psychology, deep learning has  been used to facilitate Electroencephalogram (EEG) \textcolor{black}{(the abbreviations used in this paper are summarized in Table \ref{tab:abbr})} data processing \cite{RN675}. Considering its application in material design \cite{malkiel2017deep} and quantum chemistry \cite{schutt2017quantum}, people also believe deep learning will become a valuable tool in computational chemistry \cite{goh2017deep}. Within the computational physics field, deep learning has been shown to be able to accelerate flash calculation \cite{li2018accelerating}. 


Meanwhile, deep learning has clearly demonstrated its power in promoting the bioinformatics field \cite{RN699,RN700,ching2018opportunities}, including sequence analysis \cite{zhou2015predicting,RN612,RN730,tran2017novo,RN807,angermueller2017deepcpg,RN590,wang2018wavenano,umarov2018promid,umarov2019promoter,deerect}, structure prediction and reconstruction \cite{RN661,DLBI,fout2017protein,wang2018predmp,xiong2017deep,zhang2015deep,wang2017accurate}, biomolecular property and function prediction \cite{RN775,zou2018mldeepre,kulmanov2017deepgo,almagro2017deeploc}, biomedical image processing and diagnosis \cite{RN773,kermany2018identifying,godinez2017multi,christiansen2018silico,van2016deep}, and biomolecule interaction prediction and systems biology \cite{ma2018using,zitnik2018modeling,zong2017deep,onto2vec,opa2vec,kim2018riddle,deodti,zeng2016convolutional}. Specifically, regarding sequence analysis, people have used deep learning to predict the effect of noncoding sequence variants \cite{zhou2015predicting,wang2018define}, model the transcription factor binding affinity landscape \cite{RN730,RN612,wang2018define}, improve DNA sequencing \cite{RN807,teng2018chiron} and peptide sequencing \cite{tran2017novo}, analyze DNA sequence modification \cite{RN772}, and model various post-transcription regulation events, such as alternative polyadenylation \cite{leung2017inference}, alternative splicing \cite{RN590}, transcription starting site \cite{umarov2017recognition,umarov2018promid}, noncoding RNA \cite{yang2018lncadeep,baek2018lncrnanet} and transcript boundaries \cite{shao2017deepbound}. In terms of structure prediction, \cite{RN661,RN618} use deep learning to predict the protein secondary structure; \cite{fout2017protein,zhang2015deep} adopt deep learning to model the protein structure when it interacts with other molecules; \cite{wang2017accurate,xiong2017deep,wang2018predmp} utilize deep neural networks to predict protein contact maps and the structure of membrane proteins; \cite{DLBI} accelerates the fluorescence microscopy super-resolution by combining deep learning with Bayesian inference. Regarding the biomolecular property and function prediction, \cite{RN775} predicts enzyme detailed function by predicting the Enzyme Commission number (EC numbers) using deep learning; \cite{kulmanov2017deepgo} deploys deep learning to predict the protein Gene Ontology (GO); \cite{almagro2017deeploc} predicts the protein subcellular location with deep learning. There are also a number of breakthroughs in using deep learning to perform biomedical image processing and biomedical diagnosis. For example, \cite{RN773} proposes a method based on deep neural networks, which can reach dermatologist-level performance in classifying skin cancer; \cite{kermany2018identifying} uses transfer learning to solve the data-hungry problem to promote the automatic medical diagnosis; \cite{christiansen2018silico} proposes a deep learning method to automatically predict fluorescent labels from transmitted-light images of unlabeled biological samples; \cite{godinez2017multi,van2016deep} also propose deep learning methods to analyze the cell imagining data. As for the final main direction, i.e., biomolecule interaction prediction and systems biology, \cite{ma2018using} uses deep learning to model the hierarchical structure and the function of the whole cell; \cite{zong2017deep,deodti} adopt deep learning to predict novel drug-target interaction; \cite{zitnik2018modeling} takes advantage of multi-modal graph convolutional networks to model polypharmacy sides effects.

\begin{table*}[!t]
\textcolor{black}
{\caption{\textcolor{black}{The applications of deep learning in bioinformatics.}}
\centering
\label{tab:applications}
\begin{tabular}{ |p{3cm}|p{5cm}|p{1.5cm}|p{2cm}|p{2cm}|}
 \hline
 Research direction &  Data & Data types & Candidate models & References \\
 \hline
 Sequence analysis &  Sequence data (DNA sequence, RNA sequence, \textit{et al.}) & 1D data & CNN, RNN & \cite{zhou2015predicting,RN612,RN730,tran2017novo,RN807,angermueller2017deepcpg,RN590,umarov2018promid,umarov2019promoter,deerect,wang2018wavenano} \\
 \hline
 Structure prediction and reconstruction  & MRI images, Cryo-EM images, fluorescence microscopy images, protein contact map & 2D data & CNN, GAN, VAE & \cite{RN661,DLBI,fout2017protein,wang2018predmp,xiong2017deep,zhang2015deep,wang2017accurate} \\
 \hline
 Biomolecular property and function prediction & Sequencing data, PSSM, structure properties, microarray gene expression & 1D data, 2D data, structured data & DNN, CNN, RNN & \cite{RN775,zou2018mldeepre,kulmanov2017deepgo,almagro2017deeploc} \\
 \hline
 Biomedical image processing and diagnosis & CT images, PET images, MRI images & 2D data & CNN, GAN & \cite{RN773,kermany2018identifying,godinez2017multi,christiansen2018silico,van2016deep} \\
  \hline
 Biomolecule interaction prediction and systems biology &  Microarray gene expression, PPI, gene-disease interaction, disease-disease similarity network, disease-variant network & 1D data, 2D data, structured data, graph data & CNN, GCN & \cite{ma2018using,zitnik2018modeling,zong2017deep,deodti,kordopati2018mutation,zeng2016convolutional,kim2018riddle,li2019pgcn}\\
 \hline
\end{tabular}
}
\end{table*}

In addition to the increasing computational capacity and the improved algorithms \cite{RN635,RN586,RN696,RN691,li2018decision,RN727}, the core reason for deep learning's success in bioinformatics is the data. The enormous amount of data being generated in the biological field, which was once thought to be a big challenge \cite{RN561}, actually makes deep learning very suitable for biological analysis. In particular, deep learning has shown its superiority in dealing with the following biological data types. Firstly, deep learning has been successful in handling sequence data, such as DNA sequences \cite{zhou2015predicting,wang2018define,RN772}, RNA sequences \cite{pan2018predicting,yang2018lncadeep,baek2018lncrnanet}, protein sequences \cite{RN775,kulmanov2017deepgo,almagro2017deeploc,tran2017novo}, and Nanopore signal \cite{RN807}. Trained with backpropagation and stochastic gradient descent, deep learning is expert in detecting and identifying the known and previously unknown motifs, patterns and domains hidden in the sequence data \cite{shrikumar2017learning,RN730,RN612,RN775}. Recurrent neural networks and convolutional neural networks with 1D filters are suitable for dealing with this kind of data. However, since it is not easy to explain and visualize the pattern discovered by recurrent neural networks, convolutional neural networks are usually the best choice for biological sequence data if one wants to figure out the hidden patterns discovered by the neural network \cite{shrikumar2017learning,RN612}. Secondly, deep learning is especially powerful in processing 2D and tensor-like data, such as biomedical images \cite{DLBI,RN773,christiansen2018silico,godinez2017multi,van2016deep} and gene expression profiles \cite{way2017extracting,chen2016gene}. The standard convolutional neural networks \cite{RN414} and their variants, such as residual networks \cite{RN680}, densely connected networks \cite{RN691}, and dual path networks \cite{RN770}, have shown impressive performance in dealing with biomedical data \cite{christiansen2018silico,godinez2017multi}. With the help of convolutional layers and pooling layers, these networks can systematically examine the patterns hidden in the original map in different scales and map the original input to an automatically determined hidden space, where the high level representation is very informative and suitable for supervised learning. Thirdly, deep learning can also be used to deal with graph data \cite{kipf2016semi,zitnik2017predicting,zitnik2018modeling,li2019pgcn}, such as symptom-disease networks \cite{zhou2014human}, gene co-expression networks \cite{yang2014gene}, protein-protein interaction networks \cite{rual2005towards} and cell system hierarchy \cite{ma2018using}, and promote the state-of-the-art performance \cite{zitnik2017predicting,zitnik2018modeling}. The core task of handling networks is to perform node embedding \cite{hamilton2017representation}, which can be used to perform downstream analysis, such as node classification, interaction prediction and community detection. Compared to shallow embedding, deep learning based embedding, which aggregates the information for the node neighbors in a tree manner, has less parameters and is able to incorporate domain knowledge \cite{hamilton2017representation}. It can be seen that all the aforementioned three kinds of data are raw data without much feature extraction process when we input the data to the model. Deep learning is very good at handling the raw data since it can perform feature extraction and classification in an end-to-end manner, which determines the important high level features automatically. As for the structured data which have already gone through the feature extraction process, deep learning may not improve the performance significantly. However, it will not be worse than the conventional methods, such as support vector machine (SVM), as long as the hyper-parameters are carefully tunned. \textcolor{black}{The applications of deep learning in those research directions and datasets are summarized in Table \ref{tab:applications}. }

Considering the great potential of deep learning in promoting the bioinformatic research, to facilitate the the development and application, in this review, we will first give a detailed and thorough introduction of deep learning (Section \ref{sec:tutorial}), from shallow neural networks to deep neural networks and their variants mentioned above, which are suitable for biological data analysis. After that, we provide a number of concrete examples (Section \ref{sec:examples}) with implementation available on Github, covering five bioinformatic research directions (sequence analysis, structure prediction and reconstruction, biomolecule property and function prediction, biomedical image processing and diagnosis, and biomolecular interaction prediction and systems biology) and all the four types of data (1D sequences, 2D images and profiles, graphs, and preprocessed data). In terms of network types, those examples will cover fully connected neural networks, standard convolutional neural networks (CNN) \cite{RN414}, recurrent neural networks (RNN) \cite{mikolov2010recurrent}, residual networks (ResNet) \cite{RN680}, generative adversarial networks (GAN) \cite{goodfellow2014generative}, variational autoencoder (VAE) \cite{doersch2016tutorial}, and graph convolutional neural networks (GCN) \cite{kipf2016semi}. After those concrete examples, we will discuss the potential issues which researchers may encounter when using deep learning and the corresponding possible solutions (Section \ref{sec:discussion}), including overfitting (Setion \ref{sub:overfitting}), data issues (Section \ref{sub:lacking_of_data} and \ref{sub:unbalanced_data}), interpretability (Section \ref{sub:interpretability}), uncertainty scaling (Section \ref{sub:scaling}), catastrophic forgetting (Section \ref{sub:catastrophic_forgetting}) and model compression (Section \ref{sub:reducing_computation_requirement_and_model_compression}).


\begin{table*}[!th]
\textcolor{black}
{\caption{\textcolor{black}{Summaries of related review papers.}}
\centering
\label{tab:review}
\begin{tabular}{ p{1.5cm}p{3cm}p{9cm}}
 \hline
 Reference &  Key words & Summary \\
 \hline
 \cite{leung2016machine} & Machine learning, Genomic medicine  & A review of machine learning tasks and related datasets in genomic medicine, not specific to deep learning \\
 \hline
 \cite{RN700}  & Regulatory genomics, Cellular imaging & A review of deep learning's applications in computational biology, with an emphasis on regulatory genomics and cellular imaging \\
 \hline
 \cite{RN699} & Omics, Biomedical imaging, Biomedical signal processing &  An general review of the CNN and RNN architectures, with the applications in omics, image processing and signal processing\\
 \hline
 \cite{mamoshina2016applications} & Biomarker development, Transcriptomics &  It discussed the key features of deep learning and a number of applications of it in biomedical studies \\
 \hline
 \cite{litjens2017survey} & Medical imaging, CNN & It focuses on the applications of CNN in biomedical image processing, as well as the challenges in that field. \\
 \hline
 \cite{jurtz2017introduction} & Protein sequences  & A tutorial-style review of using deep learning to handle sequence data, with three application examples \\
  \hline
 \cite{ching2018opportunities} & Biomedicine, Human disease & It provides an review of the deep learning applications in biomedicine, with comprehensive discussion of the obstacles when using deep learning \\
 \hline
 \cite{wainberg2018deep} &  Biomedicine, Model flexibility & It discusses the history and advantage of deep learning, arguing the potential of deep learning in biomedicine \\
 \hline
 \cite{esteva2019guide} & Healthcare, End-to-end system & It discusses how techniques in computer vision, natural language processing, reinforcement learning can be used to build health care system. \\
 \hline
 \cite{topol2019high} & High-performance system, Artificial intelligence, Medicine & It discusses the applications of artificial intelligence into medicine, as well as the potential challenges. \\
 \hline
 \cite{cheng2017survey} & Model compression, Acceleration & A review of methods in compressing deep learning models and reducing the computational requirements of deep learning methods. \\
 \hline
 \cite{RN808} & Catastrophic forgetting, Incremental learning & A survey of incremental learning in neural network, discussing how to incorporate new information into the model \\
 \hline
 \cite{buda2018systematic} & Class imbalance  & It investigates the impact of data imbalance to deep learning model's performance, with a comparison of frequently used techniques to alleviate the problem \\
 \hline
 \cite{kukavcka2017regularization} & Overfitting, Regularization & A systematic review of regularization methods used to combat overfitting issue in deep learning \\
 \hline
 \cite{zhang2018visual} & Interpretability, Visual interpretation & It revisits the visualization of CNN representations and methods to interpret the learned representations, with an perspective on explainable artificial intelligence  \\
 \hline
 \cite{tan2018survey} & Transfer learning, Data shortage & It reviews the transfer learning methods in deep learning field, dealing with the data shortage problem \\
 \hline
 \cite{RN654} & Model interpretability & A comprehensive discussion of the interpretability of machine learning models, from the definition of interpretability to model properties \\
 \hline
\end{tabular}
}
\end{table*}

\textcolor{black}{Notice that there are a number of other review papers available, introducing the applications of deep learning in bioinformatics, biomedicine and healthcare \citep{leung2016machine,RN700,RN699,mamoshina2016applications,jurtz2017introduction,ching2018opportunities,wainberg2018deep,esteva2019guide,topol2019high}, which are summarized in Table \ref{tab:review}. Despite their comprehensive survey of recent applications of deep learning methods to the bioinformatics field and their insightful points about the future research direction combining deep learning and biology, seldom have those reviews introduced how those algorithms work step-by-step or provided tutorial type of reviews to bridge the gap between the method developers and the end users in biology. Biologists, who do not have strong machine learning background and want incorporate the deep learning method into their data analysis pipelines, may want to know exactly how those algorithms work, avoiding the pitfalls that people usually encounter when adopting deep learning in data analytics. Thus, complementary to the previous reviews, in this paper, we provide both the exoteric introduction of deep learning, and concrete examples and implementations of its representative applications in bioinformatics, to facilitate the adaptation of deep learning into biological data analysis pipeline. This review is in tutorial-style. For the completeness of this review, we also surveyed the related works in recent years, as shown in Table \ref{tab:applications}, as well as the obstacles and corresponding solutions when using deep learning (Section \ref{sec:discussion}).}

\section{From shallow neural networks to deep learning}
\label{sec:tutorial}
In this section, we will first introduce the form of a shallow neural network and its core components (Section \ref{sub:shallow_neural_networks_and_its_components}). After that, we introduce the key components of the standard CNN and RNN (Section \ref{sub:legend_deep_learning_architectures_cnn_and_rnn}). Since the standard CNN and RNN have been improved greatly over the past few years, we will also introduce several state-of-the-art architectures (Section \ref{sub:state_of_the_art_deep_architectures}), including ResNet, DenseNet and SENet \cite{RN728}. After introducing the architecture for regular 1D and 2D data, we introduce graph neural networks for handling network data (Section \ref{sub:graph_neural_networks}). Then, we  introduce two important generative models (Section \ref{sub:generative_model_gan_and_vae}), GAN and VAE, which can be useful for biomedical image processing and drug design. Finally, we give an overview of the currently available frameworks, which make the application of deep learning quite handy, for building deep learning models (Section \ref{sub:frameworks}). \textcolor{black}{Notice that all the architectures and models are further illustrated and supported by the examples in Section \ref{sec:examples}. One can find the link between Section \ref{sec:tutorial} and Section \ref{sec:examples} with Table \ref{tab:example_sum}.}

\subsection{Shallow neural networks and their components} 
\label{sub:shallow_neural_networks_and_its_components}

\begin{figure}
  \centerline{\includegraphics[width=140mm]{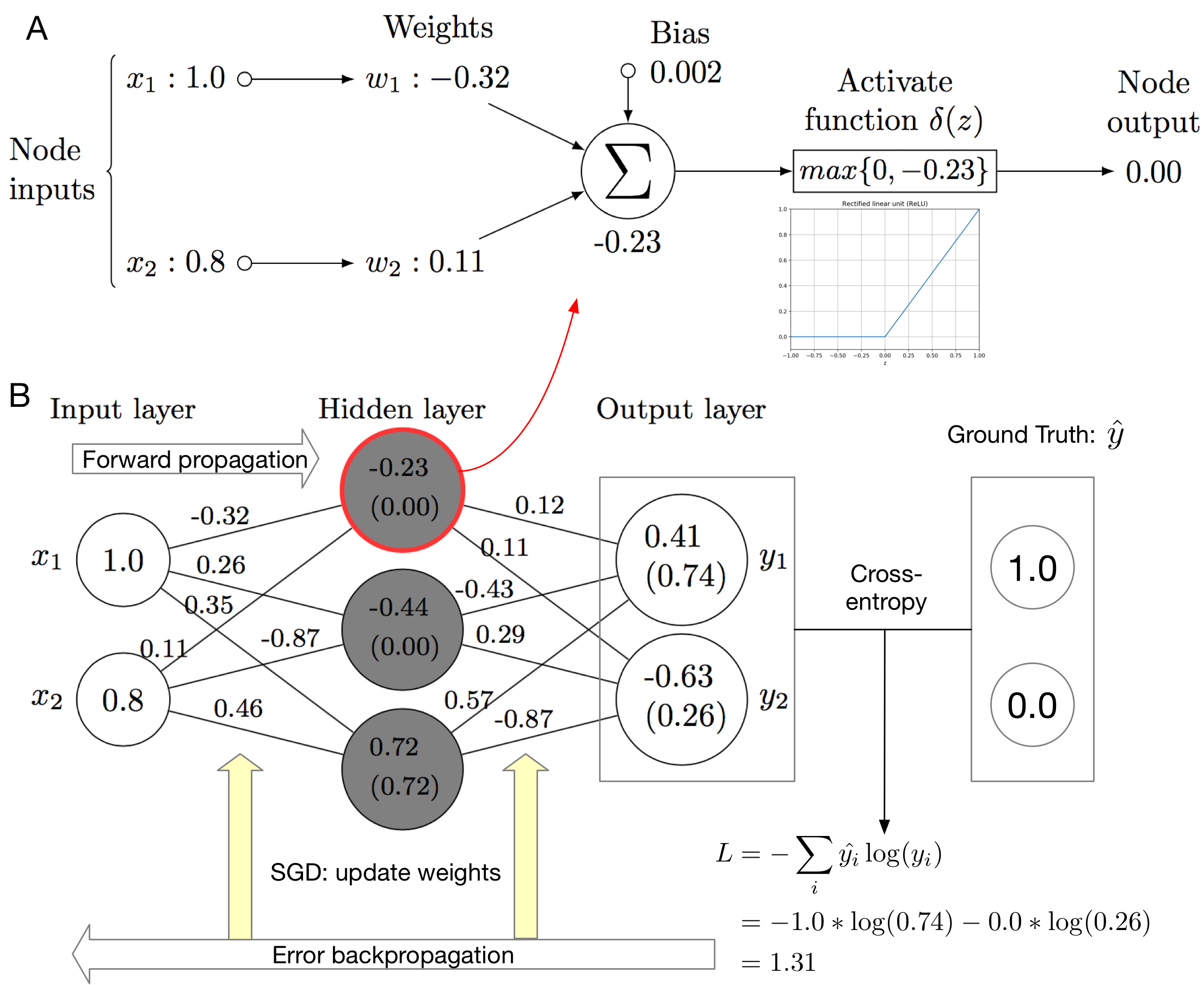}}
  \caption{Illustration of a shallow neural network. (A) The operation within a single node. The input of the node will go through a linear transformation (dot product with the weight parameters and added with a bias) and a non-linear activation function to get the node output. (B) Illustration of the key components of a shallow neural network with one hidden layer. When training the model, we will run both the forward propagation and error back-propagation. Feeding the input value to the neural network and running forward propagation, we can obtain the output of the model, with which we can compare the target value and get the difference (loss or error). Then, we run error back-propagation and optimization to adjust the parameters of the model, making the output as close to the target value as possible. When the model is well-trained and we want to use the model, we will feed the input to the model and run forward propagation to obtain the output. Notice that in the hidden layer, the activation function is ReLU and the last output layer's activation function is Softmax. The numbers in the parenthesis are the results of applying activation functions to the linear transformation outputs.}
  \label{fig:shallow_network}
\end{figure}

Fig. \ref{fig:shallow_network} shows the major components of a shallow neural network. In general, the whole network is a mapping function. Each subnetwork is also a mapping function. Taking the first hidden node as an example, which is shown in Fig. \ref{fig:shallow_network} (A), this building block function mapping contains two components, a linear transformation $\textbf{w*x}$ and a non-linear one $\delta(z)$. By aggregating multiple building block functions into one layer and stacking two layers, we obtain the network shown in Fig. \ref{fig:shallow_network} (B), which is capable of expressing a nonlinear mapping function with a relatively high complexity.

When we train a neural network to perform classification, we are training the neural network into a certain non-linear function to express (at least approximately) the hidden relationship between the features $\textbf{x}$ and the label $\hat{\textbf{y}}$. In order to do so, we need to train the parameters $\textbf{w}$, making the model fit the data. The standard algorithm to do so is forward-backward propagation. After initializing the network parameters randomly, we run the network to obtain the network output (forward propagation). By comparing the output of the model and the ground truth label, we can compute the difference (`loss' or `error') between the two with a certain loss function. Using the gradient chain rule, we can back-propagate the loss to each parameter and update the parameter with certain update rule (optimizer). We will run the above steps iteratively (except for the random initialization) until the model converges or reaches a pre-defined number of iterations. After obtaining a well-trained neural network model, when we perform testing or use it, we will only run the forward propagation to obtain the model output given the input.

From the above description, we can conclude that multiple factors in different scales can influence the model's performance. Starting from the building block shown in Fig. \ref{fig:shallow_network} (A), the choice of the activation function can influence the model's capacity and the optimization steps greatly. Currently, the most commonly used activation functions are ReLU, Leaky ReLU, SELU, Softmax, TanH. Among them, ReLU is the most popular one for the hidden layers and the Softmax is the most popular one for the output layer. When combining those building blocks into a neural network, we need to determine how many blocks we want to aggregate in one layer. The more nodes we put into the neural network, the higher complexity the model will have. If we do not put enough nodes, the model will be too simple to express the complex relationship between the input and the output, which results in underfitting. If we put too many nodes, the model will be so complex that it will even express the noise in the data, which causes overfitting. We need to find the balance when choosing the number of nodes. After building the model, when we start training the model, we need to determine which loss function we want to use and which optimizer we prefer. The commonly used loss functions are cross-entropy for classification and mean squared error for regression. The optimizers include stochastic gradient descent (SGD), Momentum, Adam \cite{RN598} and RMSprop. Typically, if users are not very familiar with the problem, Adam is recommended. If users have clear understanding of the problem, Momentum with learning rate decay is a better option.

\subsection{Legendary deep learning architectures: CNN and RNN} 
\label{sub:legend_deep_learning_architectures_cnn_and_rnn}

\subsubsection{Legendary convolutional neural networks} 
\label{ssub:legend_convolutional_neural_network}

\begin{figure}
  \centerline{\includegraphics[width=150mm]{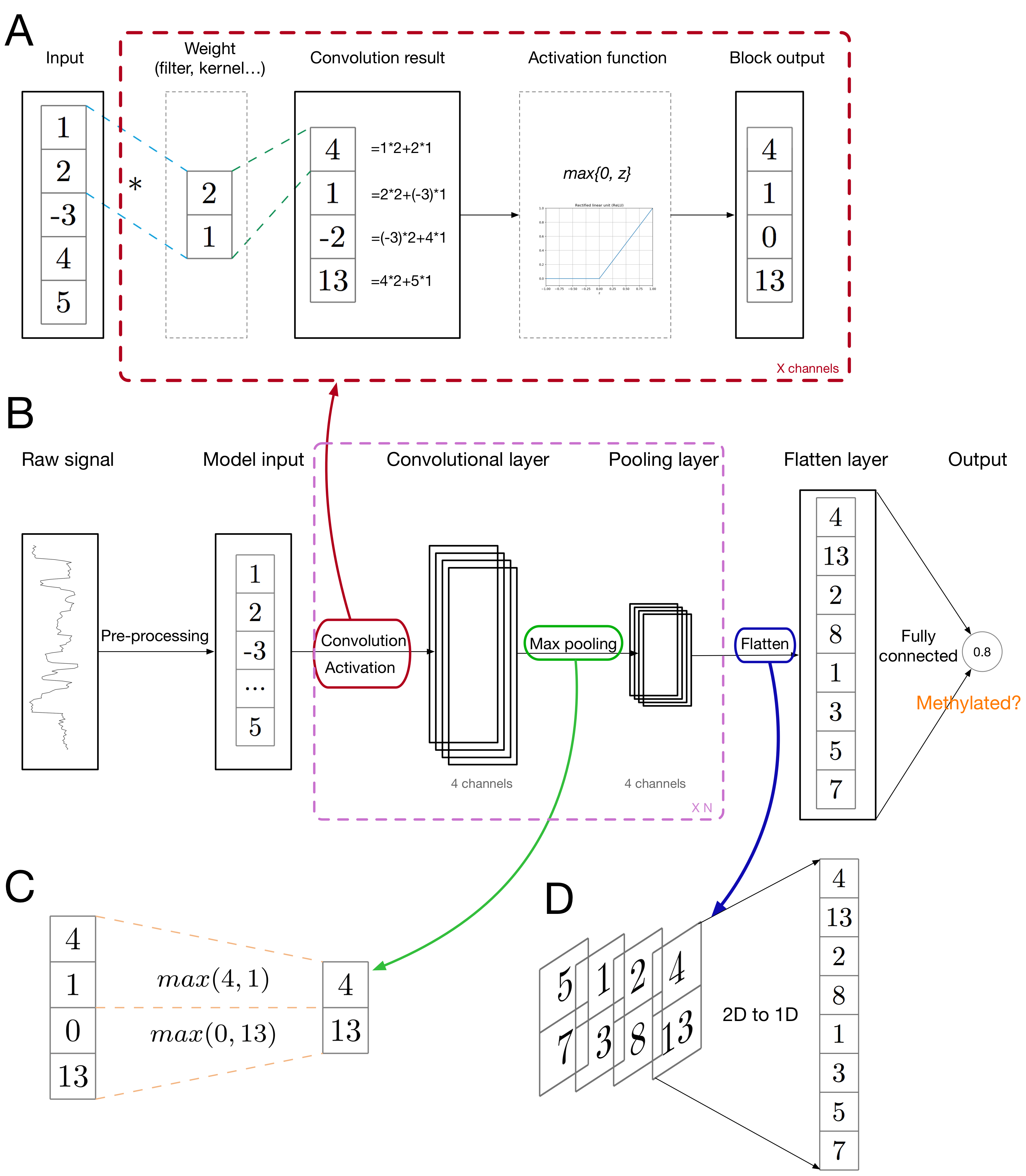}}
  \caption{Illustration of a convolution neural network. (A) The convolutional layer operation. The input goes through a convolution operation and a non-linear activation operation to obtain the output. Those two operations can be repeated $X$ times, so the output can have $X$ channels. (B) The general data flow of a convolutional neural network. After pre-processing, the data usually go through convolutional layers, pooling layers, flatten layers, and fully connected layers to produce the final output. As for the other components, such as back-propagation and optimization, it is the same as the shallow neural network. (C) The max pooling operation. (D) The flatten operation, which converts a vector with multiple channels into a long vector with only one channel.}
  \label{fig:convolution}
\end{figure}

Although the logic behind the shallow neural network is very clear, there are two drawbacks of shallow neural networks. Firstly, the number of parameters is enormous: consider one hidden layer having $N_1$ nodes and the subsequent output layer containing $N_2$ nodes, then, we will have $N_1*N_2$ parameters between those two layers. Such a large number of parameters can cause serious overfitting issue and slow down the training and testing process. Secondly, the shallow neural network considers each input feature independently, ignoring the correlation between input features, which is actually common in biological data. For example, in a DNA sequence, a certain motif may be very important for the function of that sequence \cite{RN612}. Using the shallow network, however, we will consider each base within the motif independently, instead of the motif as a whole. Although the large amount of parameters may have the capability of capturing that information implicitly, a better option is to incorporate that explicitly. To achieve that, the convolution neural network has been proposed, which has two characteristics: local connectivity and weight sharing. We show the structure of a legendary convolutional neural network in Fig. \ref{fig:convolution}. Fig. \ref{fig:convolution} (B) shows the data flow logic of a typical convolutional neural network, taking a chunk of Nanopore raw signals as input and predicting whether the corresponding DNA sequence is methylated or not. The Nanopore raw signals are already 1D numerical vectors, which are suitable for being fed into a neural network (in terms of string inputs, such as DNA sequences and protein sequences in the fasta format, which are common in bioinformatics, the encoding will be discussed in Section \ref{sec:examples}). After necessary pre-processing, such as denoising and normalization, 
the data vector will go through several ($N$) convolutional layers and pooling layers, after which the length of the vector becomes shorter but the number of channels increases (different channels can be considered to represent the input sequence from different aspects, like the RGB channels for an image). After the last pooling layer, the multi-channel vector will be flatten into a single-channel long vector. The long vector fully connects with the output layer with a similar architecture shown in Fig. \ref{fig:shallow_network}.

Fig. \ref{fig:convolution} (A) shows the convolutional layer within CNN in more details. As shown in the figure, for each channel of the convolutional layer output, we have one weight vector. The length of this vector is shorter than the input vector. The weight vector slides across the input vector, performing inner product at each position and obtaining the convolution result. Then, an activation function is applied to the convolution result elementwisely, resulting in the convolutional layer output. Notice that each element of the output is only related to a certain part of the input vector, which is referred as local connectivity. Besides, unlike the structure in Fig. \ref{fig:shallow_network}, in convolutional layers, different parts of the input vector share the same weight vector when performing the sliding inner product (convolution), which is the weight-sharing property. Under this weight-sharing and local connectivity setting, we can find that the weight vector serves as a pattern or motif finder, which satisfies our original motivation of introducing this architecture. Notice that the original input is a vector with one channel. After the first convolutional layer, the input of the next convolutional layer usually has more than one channel. Correspondingly, the weight vector should have the same number of channels to make the convolutional operation feasible.  For example, in Fig. \ref{fig:convolution} (B), the output of the first convolutional layer has 4 channels. Accordingly, the filters of the next convolutional layer should have 4 channels. If the filter length is the same as what is shown in Fig. \ref{fig:convolution} (A), which is 2, the filters should have the dimensionality as 2 by 4. If the output of the second convolutional layer has 16 channels, we should have 16 filters, one for each output channel. Considering all the above factors, the filter tensor of the second convolutional layer should have a dimensionality of 2*4*16, which is a 3D tensor. For image inputs, we usually have a 4D tensor as the filter of each convolutional layer, whose four dimensions correspond to filter length, filter width, input channels and output channels, respectively.

Fig. \ref{fig:convolution} (C) shows the max pooling operation. In the max pooling layer, each element of the output is the maximum of the corresponding region in the layer input. Depending on the applications, people can choose the pooling region size and steps (filter size and stride), or a different pooling method, such as average pooling. This pooling layer enables the network to capture higher level and long range property of the input vector, such as the long range interaction between the bases within the corresponding DNA sequences of the input signals. Fig. \ref{fig:convolution} (D) shows the flatten layer. This operation is straightforward, just concatenating the input vectors with multiple channels into a long vector with only one channel to enable the downstream fully connected layer operation.

With the convolutional layers and the pooling layers, this neural network architecture is able to capture the inputs' property in different levels and at different scales, including both the local motif and long range interaction.


\subsubsection{Legendary recurrent neural networks} 
\label{ssub:legend_recurrent_neural_network}

\begin{figure}
  \centerline{\includegraphics[width=160mm]{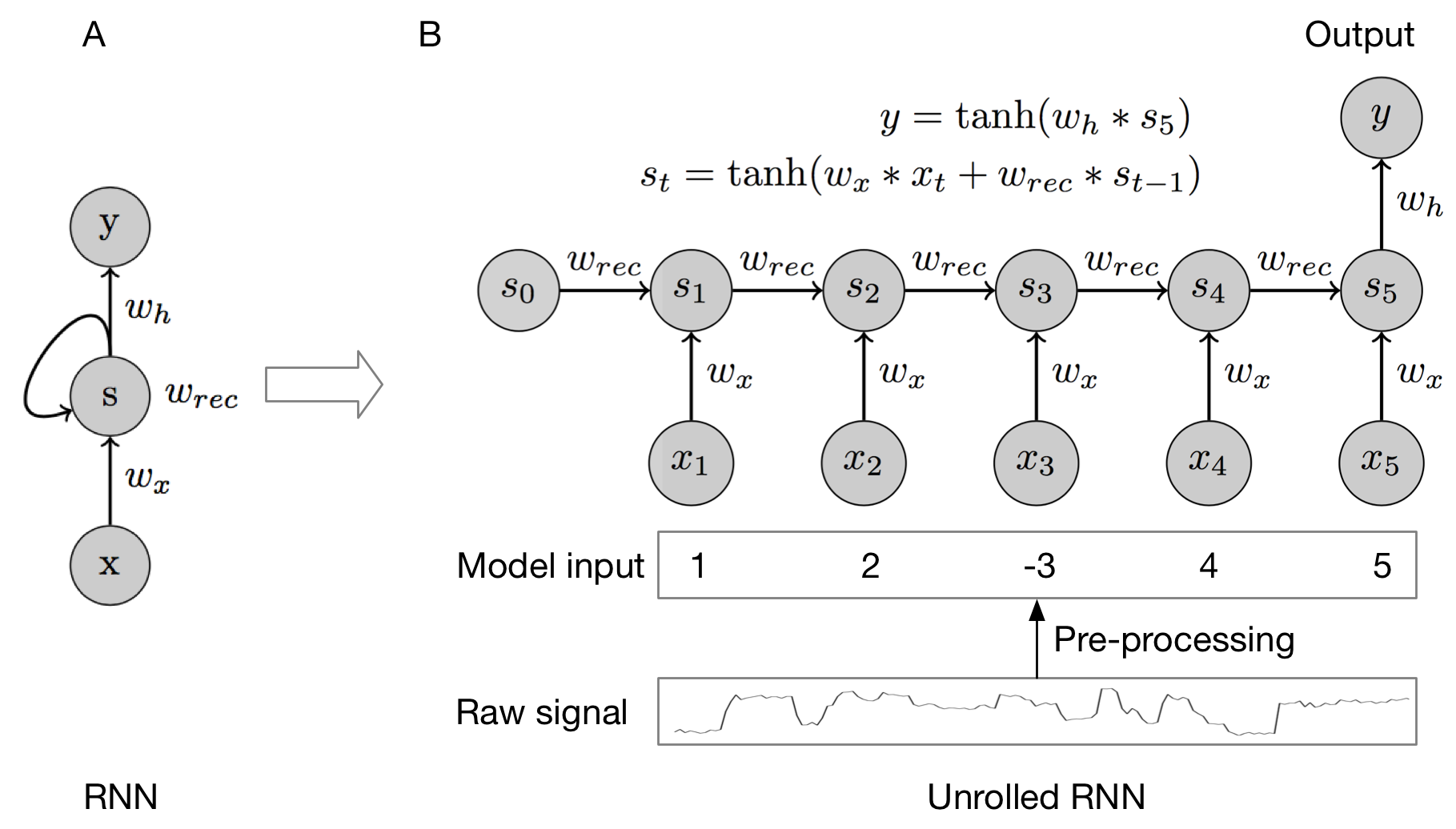}}
  \caption{Illustration of a recurrent neural network. (A) A typical rolled RNN representation. (B) An easy-to-understand unrolled RNN representation. The model processes the input element by element which enables the model to capture the temporal information within the input.}
  \label{fig:recurrent}
\end{figure}

In addition to the spatial dependency within the input, we also need to consider the temporal or order dependency in the input. For example, in DNA sequences, the order of motifs can influence the function of the sequence chunk \cite{quang2016danq}. In a document, we need to consider the order of words and sentences to categorize the document \cite{liu2016recurrent}. Similar to the case of spatial information, since the shallow neural networks shown in Fig. \ref{fig:shallow_network} consider each element of the input independently, the temporal information may also be lost. Recurrent neural networks are specifically designed to exploit the temporal relationship within a sequential input. The basic structure of a recurrent neural network is shown in Fig. \ref{fig:recurrent}. Let us consider an RNN with only one hidden recurrent cell, as shown in Fig. \ref{fig:recurrent} (A). To explain how RNN works, we unroll the network into Fig. \ref{fig:recurrent} (B), using the simplified Nanopore raw signals as an example. As in CNN, we first perform necessary pre-processing, such as normalization and denoising. Then the vector will be fed to the model element by element. The hidden recurrent cell has an initial state, which is denoted as $s_0$ and can be initialized randomly. After taking the first value, the recurrent node is updated, considering the previous state $s_0$ and $x_1$. Under our setting, the update rule is $s_1 = \tanh(w_x*x_1+w_{rec}*s_0)$. When the second value comes in, the node state is updated to $s_2$ with the same rule. We repeat the above process until we consider all the elements. Then the node information will be fed forward to make predictions, e.g., whether the corresponding DNA sequence of the input Nanopore raw signals is methylated or not, with the similar downstream structure in Fig. \ref{fig:shallow_network}. Notice that $w_x$ and $w_{rec}$ are shared among all the time steps or positions for that specific hidden recurrent node. Usually, one single recurrent node is not enough to capture all the temporal information. Under that circumstance, we can increase the number of recurrent nodes, each with its own pair of $w_x$ and $w_{rec}$. For the last hidden layer, we can use a fully connected layer to connect all the hidden recurrent nodes to the output node, like we is done in the CNN scenario. In terms of training and optimization, it is the same as the case in shallow neural networks. We will run error back-propagation to make the weight parameters fit the training data and perform prediction afterwards.

\subsection{State-of-the-art deep architectures} 
\label{sub:state_of_the_art_deep_architectures}

\begin{figure}
  \centerline{\includegraphics[width=180mm]{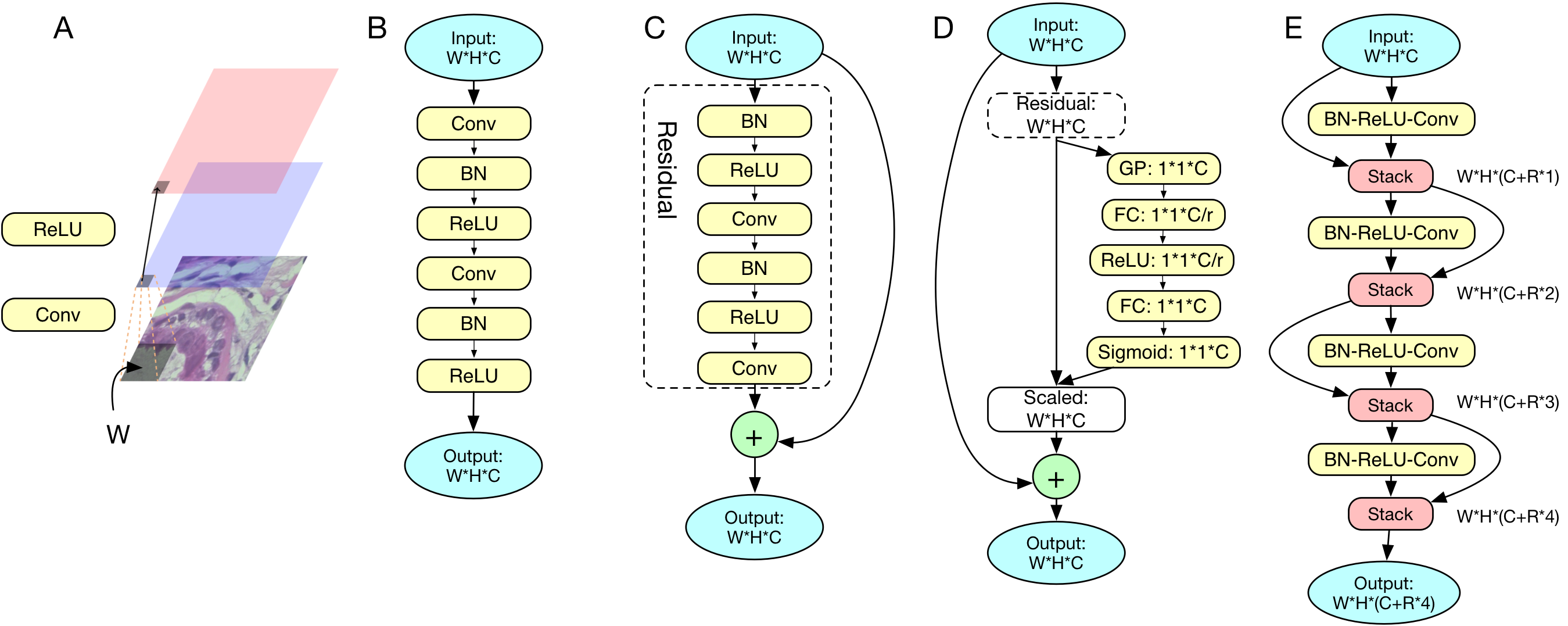}}
  \caption{Illustration of different convolutional neural network architectures. (A) The convolutional neural network building block. `Conv' represents the convolutional operation. `ReLU' represents the activation function. (B) The legendary convolutional neural network with two convolutional layers. Each convolutional layer contains `Conv', `BN', and `ReLU', where `BN' represents batch normalization, which accelerates the training process \cite{RN635}. (C) The residual block of a residual network \cite{RN680}. In addition to the legendary convolutional layers along the data flow process, there is a shortcut between the input and the output, which adds the input elementwisely to the convolutional layer output to obtain the residual block output. In fact, unlike the architecture in (B), where the convolutional layers are used to model the mapping between the input and the output, here in (C), those convolutional layers are used to model the residual between the input and the residual block output. (D) The building block of SENet \cite{RN728}. This architecture further improves the residual block, which introduces the idea of Attention \cite{RN788} into the residual network. In (C), different channels of the output are considered of equal importance. However, in reality, some channels can contain more information and thus more important. The SENet block considers that, learning a scaling function and applying the scaling function to the output of the residual block to obtain the SENet block output. In the figure, `GP' represents global pooling; `FC' represents a fully connected layer. (E) The DenseNet block \cite{RN691}. Unlike (C,D), which have the elementwise addition shortcut, DenseNet block's shortcut is stacking operation (`stack' in the figure), which concatenates the original input to the output of a certain convolutional layer along the channel dimension. This block can result in outputs with a large number of channels, which is proved to be a better architecture.}
  \label{fig:architecture}
\end{figure}

In the past several years, the above two legendary deep learning architectures have been improved greatly. For convolutional neural networks, AlexNet \cite{RN414}, VGG \cite{RN567}, GoogleNet \cite{RN620,RN676}, ResNet \cite{RN696,RN680}, ResNext \cite{RN706}, SENet \cite{RN728}, DenseNet \cite{RN691} and DPN \cite{RN770} have been proposed. For recurrent neural networks, there are LSTM \cite{gers1999learning}, Bi-RNN \cite{graves2013hybrid}, GRU \cite{chung2014empirical}, Memory network \cite{weston2014memory} and Attention network \cite{RN788}. In Fig. \ref{fig:architecture}, we exhibit some typical CNN architectures to deal with 2D image data, showing the evolving of convolutional neural networks. In general, the more advanced CNN architectures allow people to stack more layers, with the hope to extract more useful hidden informations from the input data at the expense of more computational resources. For recurrent neural networks, the more advanced architectures help people deal with the gradient vanishing or explosion issue and accelerate the execution of RNN. However, how to parallelize RNN is still a big problem under active investigation \cite{yu2018sliced}.

\subsection{Graph neural networks} 
\label{sub:graph_neural_networks}
\begin{figure}
  \centerline{\includegraphics[width=160mm]{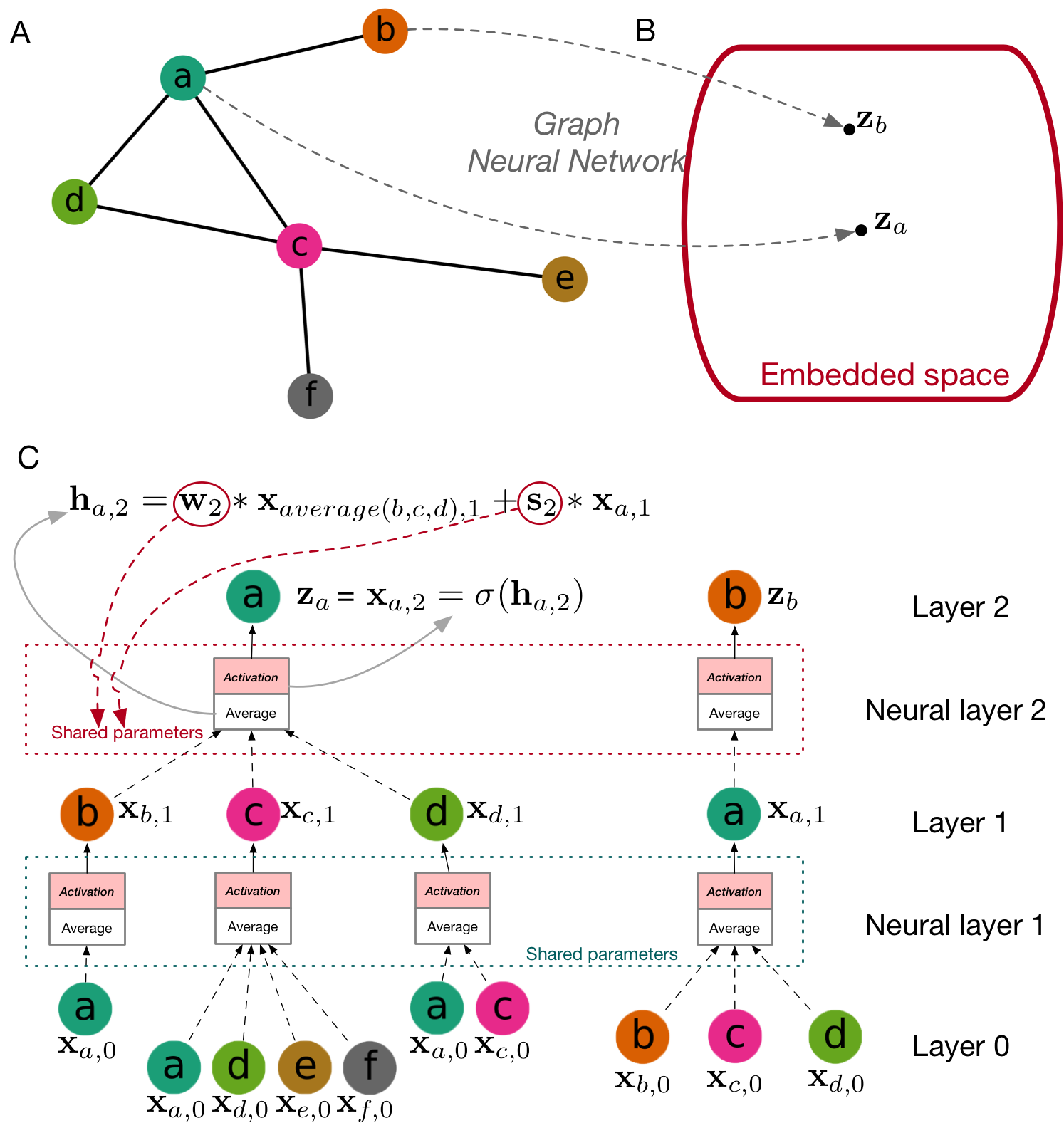}}
  \caption{Illustration of a graph neural network. (A) A typical example of graph data. (B) The embedding space. In this embedding space, each data point is represented by a vector while the original topological information in (A) is preserved in that vector. (C) The graph neural network for embedding the network in (A). We use node \textit{a} and \textit{b} as examples. The internal properties of each node are considered as the original representations. In each layer, the nodes aggregate information from their neighbors and update the representations with averaging and activation function.  The output of layer 2 are considered as the embedding result in this example. Notice that the parameters within the same layer between different trees are shared so this method can be generalized to previously unseen graph of the same type.}
  \label{fig:gnn}
\end{figure}

In this section, we briefly introduce graph neural networks \cite{kipf2016semi,hamilton2017representation} to deal with network data, which is a common data type in bioinformatics. Unlike the sequence and image data, network data are irregular: a node can have arbitrary connection with other nodes. Although the data are complex, the network information is very important for bioinformatics analysis, because the topological information and the interaction information often have a clear biological meaning, which is a helpful feature for perform classification or prediction. When we deal with the network data, the primary task is to extract and encode the topological and connectivity information from the network, combining that information with the internal property of the node \cite{hamilton2017representation}. For example, Fig. \ref{fig:gnn} (A) shows a protein-protein interaction network and each node in the network represents a protein. In addition to the interaction information, each protein has some internal properties, such as the sequence information and the structure information. If we want to predict whether a protein in the network is an enzyme, both the interaction information and the internal properties can be helpful. So, what we need to do is to encode the protein in the network in a way that we consider both the network information and the properties, which is known as an embedding problem. In other words, we embed the network data into a regular space, as shown in Fig. \ref{fig:gnn} (B), where the original topological information is preserved. For an embedding problem, the most important thing is to aggregate information from a node's neighbor nodes \cite{hamilton2017representation}. Graph convolutional neural networks (GCN), shown in Fig. \ref{fig:gnn} (C) are designed for such  a purpose. Suppose we consider a graph neural network with two layers. For each node, we construct a neighbor tree based on the network (Fig. \ref{fig:gnn} (C) shows the tree constructed for node \textit{a} and node \textit{b}). Then we can consider layer 0 as the neural network inputs, which can be the proteins' internal property encoding in our setting. Then, node \textit{a}'s neighbors aggregate information from their neighbors, followed by averaging and activating, to obtain their level 1 representation. After that, node \textit{a} collects information from its neighbors' level 1 representation to obtain its level 2 representation, which is the neural networks' output embedding result in this example. 

Take the last step as an example, the information collection (average) rule is: $\textbf{h}_{a,2} = \textbf{w}_2*\textbf{x}_{average(b,c,d),1}+\textbf{s}_2*\textbf{x}_{a,1}$, where $\textbf{x}_{average(b,c,d),1}$ is the average of node \textit{b,c,d}'s level 1 embedding, $\textbf{x}_{a,1}$ is node \textit{a}'s level 1 embedding, and $\textbf{w}_2$ and $\textbf{s}_2$ are the trainable parameters. To obtain the level 2 embedding, we apply an activation function to the average result: $\textbf{x}_{a,2} = \sigma(\textbf{h}_{a,2})$. Notice that between different nodes, the weights within the same neural layer are shared, which means the graph neural network can be generalized to previously unseen network of the same type. To train the graph neural network, we need to define a loss function, with which we can use the back-propagation to perform optimization. The loss function can be based on the similarity, that is, similar nodes should have similar embeddings \cite{grover2016node2vec,perozzi2014deepwalk,rhee2017hybrid}. Or we can use a classification task directly to train the GCN in a discriminative way \cite{dutil2018towards,kipf2016semi,zitnik2017predicting}: we can stack a shallow neural network, CNN or RNN, on the top of GNN, taking the embedding output of the GCN as input and training the two networks at the same time. 

\begin{figure}[t]
  \centerline{\includegraphics[width=150mm]{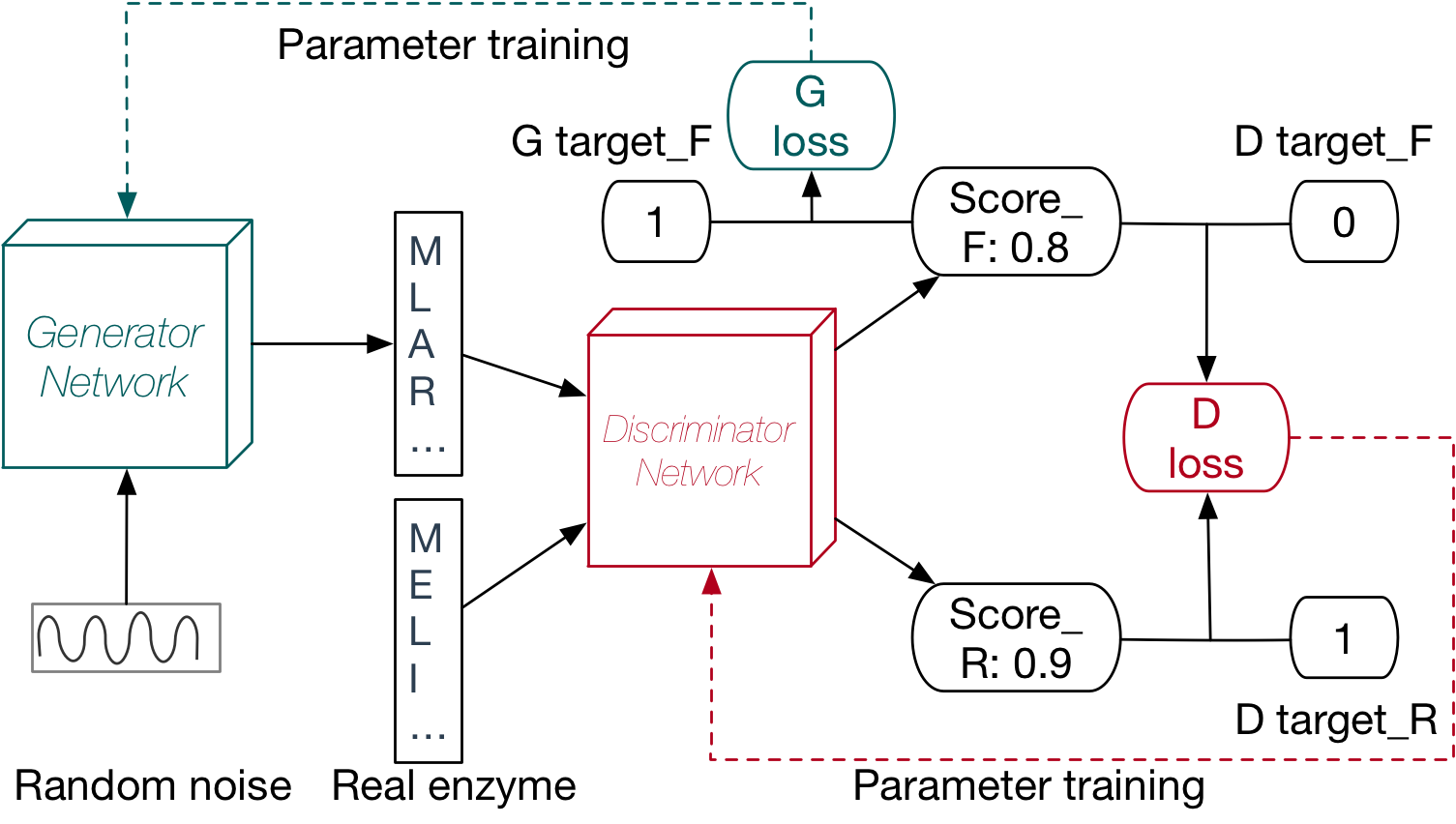}}
  \caption{Illustration of GAN. In GAN, we have a pair of networks competing with each other at the same time. The generator network is responsible for generating new data points (enzyme sequences in this example). The discriminator network tries to distinguish the generated data points from the real data points. As we train the networks, both models' abilities are improved. The ultimate goal is to make the generator network able to generate enzyme sequences that are very likely to be real ones that have not been discovered yet. The discriminator network can use the architectures in Fig. \ref{fig:architecture}. As for the generator network, the last layer should have the same dimensionality as the enzyme encoding.}
  \label{fig:gan}
\end{figure}

\subsection{Generative models: GAN and VAE} 
\label{sub:generative_model_gan_and_vae}
In this section, we introduce two generative networks, GAN \cite{goodfellow2014generative} and VAE \cite{doersch2016tutorial}, which can be useful for biological and biomedical image processing \cite{yang2018low,DLBI,seeliger2018generative} and protein or drug design \cite{sanchez2018inverse,popova2018deep,rampasek2017dr}. Unlike supervised learning, using which we perform classification or regression, such as the task of predicting whether a protein is an enzyme, the generative models belong to unsupervised learning, which cares more about the intrinsic properties of the data. With generative models, we want to learn the data distribution and generate new data points with some variations. For example, given a set of protein sequences which are enzymes as the training dataset, we want to train a generative model which can generate new protein sequences that are also enzymes.

Generative adversarial networks (GAN) have achieved great success in the computer vision field \cite{goodfellow2014generative}, such as generating new semantic images \cite{goodfellow2014generative}, image style transfer \cite{pix2pix2017,CycleGAN2017}, image inpainting \cite{yu2018generative} and image super-resolution \cite{RN666,DLBI}. As shown in Fig. \ref{fig:gan}, instead of training only one neural network, GAN trains a pair of networks which compete with each other. The generator network is the final productive neural network which can produce new data samples (novel enzyme sequences in this example) while the discriminator network distinguishes the designed enzyme sequences from the real ones to push the generator network to produce protein sequences that are more likely to be enzyme sequences instead of some random sequences. Both of the generator network and the discriminator network can be the networks mentioned in Section \ref{ssub:legend_convolutional_neural_network}. For the generator network, the last layer needs to be redesigned to match the dimensionality of an enzyme sequence encoding. 

\begin{figure}[t]
  \centerline{\includegraphics[width=140mm]{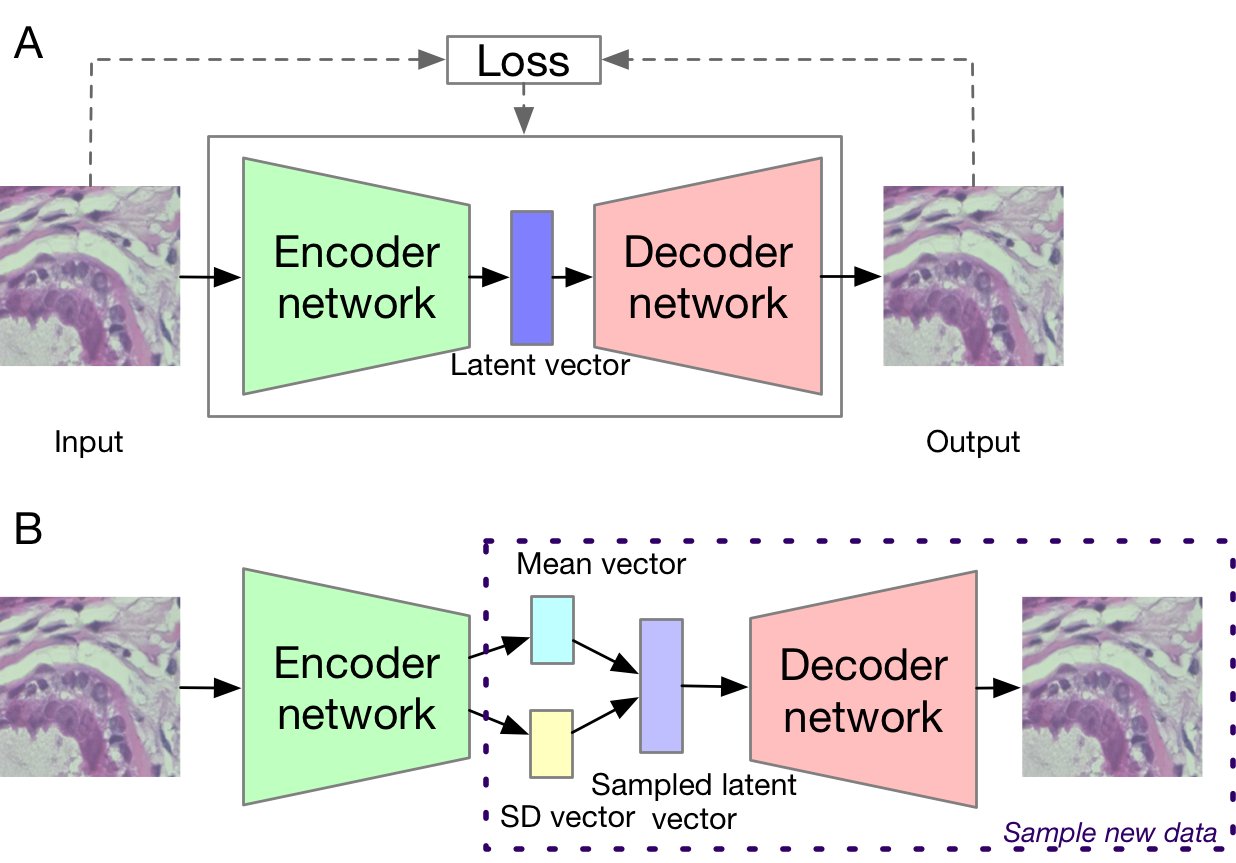}}
  \caption{(A) Illustration of autoencoder, which is composed of encoder and decoder. The encoder network compresses the input into the latent vector and the decoder network reconstructs data from the latent vector. The loss is usually defined as the difference between the input and the decoder output. Notice that the latent vector's dimensionality is much smaller than the original input. (B) Illustration of VAE. To enable the network to sample new data points, instead of mapping the inputs into fixed latent vectors, VAE maps the inputs into a distribution in the latent space with the encoder network. When we need to sample new data, we can first sample the latent vector from the distribution and then generate the data using the decoder network with the sampled latent vector as input.}
  \label{fig:vae}
\end{figure}

To introduce variational autoencoder, let us first go through autoencoder, which is shown in Fig. \ref{fig:vae} (A). The autoencoder is usually used to encode high dimensional input data, such as an image, into a much lower dimensional representation, which can store the latent information of the data distribution. It contains two parts, the encoder network and the decoder network. The encoder network can transform the input data into a latent vector and the decoder network can reconstruct the data from the latent vector with the hope that the reconstructed image is as close to the original input as possible. Autoencoder is very convenient for dimensionality reduction. However, we cannot use it to generate new data which are not in the original input. Variational antoencoder overcomes the bottleneck by making a slight change to the latent space. Instead of mapping the input data to one exact latent vector, we map the data into a low dimensional data distribution, as shown in Fig. \ref{fig:vae} (B). With the latent distribution, when we need to sample new data points, we can first sample the latent vector from the latent vector distribution and then construct the new data using the decoder network.

\subsection{Frameworks} 
\label{sub:frameworks}
Building the whole network and implementing the optimizers completely from scratch will be very tedious and time-consuming. Fortunately, there are a number of handy frameworks available, which can accelerate the process of building networks and training the model. After evolving for several years, the following frameworks are commonly used, and actively developed and maintained: Tensorflow \cite{tensorflow}, Pytorch \cite{paszke2017automatic}, Caffe2, MXNet \cite{chen2015mxnet}, CNTK \cite{seide2016cntk} and PaddlePaddle. Another famous warpper of Tensorflow is Keras, with which one can build a model in several lines of code. It is very convenient and easy-to-use, although it lacks the flexibility provided by Tensorflow, using which one can control almost every single detail.


\begin{table*}[!t]
\textcolor{black}
{\caption{\textcolor{black}{Summary of the examples.}}
\centering
\label{tab:example_sum}
\begin{tabular}{ |p{3cm}|p{1cm}|p{1.5cm}|p{4cm}|p{3cm}|}
 \hline
 Example &  Model & Data type & Research direction & Task \\
 \hline
 Enzyme function prediction &  DNN & Structured & Biomolecular function prediction & Classification \\
 \hline
 Gene expression regression  &  DNN & Structured & Biomolecular property prediction & Regression \\
 \hline
 RNA-protein binding sites prediction & CNN & 1D data & Sequence analysis & Classification \\
 \hline
 DNA sequence function prediction & CNN, RNN & 1D data & Sequence analysis & Classification \\
  \hline
 Biomedical image classification &  ResNet & 2D data & Biomedical image processing & Classification\\
 \hline
 Protein interaction prediction &  GCN & Graph & Biomolecule interaction prediction & Embedding, Classification\\
 \hline
 Biology image super-resolution &  GAN & 2D image & Structure reconstruction & Data generation\\
 \hline
 Gene expression data embedding & VAE & 2D data & Systems biology & DR, Data generation \\
 \hline
\end{tabular}
}
\end{table*}

\section{Applications of deep learning in bioinformatics}
This section provides eight examples.
\textcolor{black}{Those examples are carefully selected, typical examples of applying deep learning methods into important bioinformatic problems, which can reflect all of the above discussed research directions, models, data types, and tasks, as summarized in Table \ref{tab:example_sum}.}
In terms of research directions,
Sections \ref{sub:rna_protein_binding_sites_prediction_with_cnn} and
\ref{sub:dna_sequence_function_prediction_with_cnn_and_rnn} are related to 
sequence analysis; Section \ref{sub:biology_image_super_resolution_using_gan} is relevant to structure prediction and reconstruction; Section \ref{sub:identifying_enzyme_using_multi_layer_neural_network} is about biomolecular property and function prediction; Sections \ref{sub:biology_image_super_resolution_using_gan} and \ref{sub:biomedical_image_classification_using_transfer_learning_and_resnet} are related to biomedical image processing and diagnosis; Sections \ref{sub:gene_expression_regression}, \ref{sub:graph_embedding_for_novel_protein_interaction_prediction_using_GNN}, and \ref{sub:high_dimension_biological_data_embedding_and_generation_with_vae} are relevant to biomolecule interaction prediction and systems biology. Regarding the data type, Sections \ref{sub:identifying_enzyme_using_multi_layer_neural_network} and \ref{sub:gene_expression_regression} use structured data; Sections \ref{sub:rna_protein_binding_sites_prediction_with_cnn} and \ref{sub:dna_sequence_function_prediction_with_cnn_and_rnn} use 1D sequence data; Sections \ref{sub:biomedical_image_classification_using_transfer_learning_and_resnet},
\ref{sub:biology_image_super_resolution_using_gan}, and
\ref{sub:high_dimension_biological_data_embedding_and_generation_with_vae} use 2D image or profiling data; Section \ref{sub:graph_embedding_for_novel_protein_interaction_prediction_using_GNN} uses graph data. 

From the deep learning point of view, those examples also cover a wide range of deep learning models: Sections \ref{sub:identifying_enzyme_using_multi_layer_neural_network}  and \ref{sub:gene_expression_regression} use deep fully connected neural networks; Section \ref{sub:rna_protein_binding_sites_prediction_with_cnn} uses CNN; Section \ref{sub:dna_sequence_function_prediction_with_cnn_and_rnn} combines CNN with RNN; Section \ref{sub:graph_embedding_for_novel_protein_interaction_prediction_using_GNN} uses GCN; Section \ref{sub:biomedical_image_classification_using_transfer_learning_and_resnet} uses ResNet; Section \ref{sub:biology_image_super_resolution_using_gan} uses GAN; Section \ref{sub:high_dimension_biological_data_embedding_and_generation_with_vae} uses VAE. We cover both supervised learning, including classification (Sections \ref{sub:identifying_enzyme_using_multi_layer_neural_network}, \ref{sub:rna_protein_binding_sites_prediction_with_cnn}, \ref{sub:dna_sequence_function_prediction_with_cnn_and_rnn}, and \ref{sub:biomedical_image_classification_using_transfer_learning_and_resnet}) and regression (Section \ref{sub:gene_expression_regression}), and unsupervised learning (Sections \ref{sub:graph_embedding_for_novel_protein_interaction_prediction_using_GNN}, \ref{sub:biology_image_super_resolution_using_gan}, and \ref{sub:high_dimension_biological_data_embedding_and_generation_with_vae}). We also introduce transfer learning using deep learning briefly (Section \ref{sub:biomedical_image_classification_using_transfer_learning_and_resnet}). 


\label{sec:examples}

\subsection{Identifying enzymes using multi-layer neural networks} 
\label{sub:identifying_enzyme_using_multi_layer_neural_network}
Enzymes are one of the most important types of molecules in human body, catalyzing biochemical reactions \textit{in vivo}. Accurately identifying enzymes and predicting their function can benefit various fields, such as biomedical diagnosis and industrial bio-production \cite{RN775}. In this example, we show how to identify enzyme sequences based on sequence information using deep learning based methods.

Usually, a protein sequence is represented by a string (such as, `MLAC...'), but deep learning models, as mathematical models, take numerical values as inputs. So before building the deep learning model, we need to first encode the protein sequences into numbers. The common ways of encoding a protein sequences are discussed in \cite{RN775}. In this example, we use a sparse way to encode the protein sequences, the functional domain encoding. For each protein sequence, we use HMMER \cite{RN524} to search it against the protein functional domain database, Pfam \cite{RN652}. If a certain functional domain is hit, we encode that domain as 1, otherwise 0. Since Pfam has 16306 functional domains, we have a 16306D vector, composed of 0s and 1s, to encode each protein sequence. Because the dimensionality of the feature is very high, the traditional machine learning method may encounter the curse of dimensionality. However, the deep learning method can handle the problem quite well. As for the dataset, we use the dataset from \cite{RN775}, which contains 22168 enzyme sequences and 22168 non-enzyme protein sequences, whose sequence similarity is under 40\% within each class.

In our implementation, we adopt a similar architecture as Fig. \ref{fig:shallow_network}, but with much more nodes and layers. We use ReLU as the activation function, cross-entropy loss as the loss function, and Adam as the optimizer. We utilize dropout, batch normalization and weight decay to prevent overfitting. With the help of Keras, we build and train the model in 10 lines. Training the model on a Titan X for 2 minutes, we can reach around 94.5\% accuracy, which is very close to the state-of-the-art performance \cite{RN775}. Since bimolecular function prediction and annotation is one of the main research directions of bioinformatics, researchers can easily adopt this example and develop the applications for their own problems.

\subsection{Gene expression regression} 
\label{sub:gene_expression_regression}
Gene expression data are one of the most common and useful data types in bioinformatics, which can be used to reflect the cellular changes corresponding to different physical and chemical conditions and genetic perturbations \cite{van2002gene,chen2016gene}. However, the whole genome expression profiling can be expensive. To reduce the cost of gene profiling, realizing that different genes' expression can be highly correlated, researchers have developed an affordable method of only profiling around 1000 carefully selected landmark genes and predicting the expression of the other target genes based on computational methods and landmark gene expression \cite{duan2016l1000cds,chen2016gene}. Previously, the most commonly used method is linear regression. Recently, \cite{chen2016gene} showed that deep learning based regression can outperform the linear regression method significantly since it considered the non-linear relationship between genes' expression.

In this example, we use deep learning method to perform gene expression prediction as in \cite{chen2016gene}, showing how to perform regression using deep learning. We use the Gene Expression Omnibus (GEO) dataset from \cite{chen2016gene}, which has already gone through the standard normalization procedure. For the deep learning architecture, we use a similar structure as in Section \ref{sub:identifying_enzyme_using_multi_layer_neural_network}. However, the loss function for this regression problem is very different from the classification problem in Section \ref{sub:identifying_enzyme_using_multi_layer_neural_network}. As we discussed in Section \ref{sub:shallow_neural_networks_and_its_components}, for classification problems, we usually use cross-entropy loss, while for the regression problem, we use the mean squared error as the loss function. Besides, we also change the activation function of the last layer from Softmax to TanH for this application. Using Keras, the network can be built and trained in 10 lines. Trained on a Titan X card for 2 mins, it can outperform the linear regression method by 4.5\% on a randomly selected target gene.

Our example code can be easily used and adopted for other bioinformatics regression problems. On the other hand, if one is sure about the regression target value range, then the target range can be divided into several bins and the regression problem is converted into a classification problem. With the classification problem in hand, people can use the classification examples introduced in the other sections.

\subsection{RNA-protein binding sites prediction with CNN} 
\label{sub:rna_protein_binding_sites_prediction_with_cnn}
RNA-binding proteins (RBP) play an important role in regulating biological processes, such as gene regulation \cite{glisovic2008rna,pan2018predicting}. Understanding their behaviors, for example, their binding site, can be helpful for cuing RBP related diseases. With the advancement of high-throughput technologies, such as CLIP-seq, we can verify the RBP binding sites in batches \cite{li2013starbase}. However, despite its efficiency, those high-throughput technologies can be expensive and time-consuming. Under that circumstance, machine learning based computational methods, which are fast and affordable, can be helpful to predict the RBP binding site \cite{pan2018predicting}. 

In fact, the deep learning methods are especially suitable for this kind of problems. As we know, RBP can have sequence preference, recognizing specific motifs and local structures and binding to those points specifically \cite{ray2009rapid}. On the other hand, as we discussed in Section \ref{ssub:legend_convolutional_neural_network}, CNN are especially good at detecting special patterns in different scales. Previous studies have shown the power of CNN in finding motifs \cite{pan2018predicting,RN730}. 

In this example, we show how to predict the RBP binding site using CNN, with the data from \cite{pan2018predicting}. Specifically, the task is to predict whether a certain RBP, which is fixed for one model, can bind to a certain given RNA sequence, that is, a binary classification problem. We use one-hot encoding to convert the RNA sequence strings of `AUCG' into 2D tensors. For example, for `A', we use a vector $(1,0,0,0)$ to represent it; for `U', we use a vector $(0,1,0,0)$ to represent it. Concatenating those 1D vectors into a 2D tensor in the same order as the original sequence, we obtain the one-hot encoding for a certain RNA sequence. We use a similar architecture as the model shown in Fig. \ref{fig:convolution}. We use similar settings as the previous examples in terms of the activation function, the optimizer, the loss function, \textit{etc.}. Notice that for the one-hot encoding, we can consider it either as a 2D map with 1 channel or a 1D vector with 4 channels. Correspondingly, for the convolutional layers, we can choose either 2D convolutions or 1D convolutions. In this example, we follow the previous research setting, using 2D convolutions. The original implementation \cite{pan2018predicting} is in Pytorch, which is very lengthy, we reimplemented the idea using Keras, which builds the model in 15 lines. 

Furthermore, this example can be easily adopted for other applications with similar background. That is, there are motif or pattern preferences within the input, such as transcription starting site identification \cite{umarov2017recognition,umarov2018promid} and Poly(A)  site prediction \cite{RN553}.

\subsection{DNA sequence function prediction with CNN and RNN} 
\label{sub:dna_sequence_function_prediction_with_cnn_and_rnn}
Understanding the properties and functions of DNA sequences, which is the most important genetic material, is a profound and challenging task for bioinformatics and biology \cite{venter2001sequence}. Among those sequences, the non-coding DNA functionality determination is especially challenging, since over 98\% of human genome are non-coding DNA and it is very difficult to perform biological experiment to investigate functionality of every non-coding DNA sequence chunk \cite{pennacchio2006vivo}. Computational methods, which are cheap and highly parallelable, can help people address the problem greatly. Previous studies have shown the success of using deep learning to predict the functionality of non-coding DNA sequences \cite{zhou2015predicting,quang2016danq}.

In this example, we show how to use CNN and RNN to predict the functionality of non-coding DNA sequences. We use the data from \cite{zhou2015predicting}. As described in \cite{zhou2015predicting,quang2016danq}, the human GRCh37 reference genome was segmented into non-overlapping 200-bp bins. The inputs of the deep learning model are the 1000-bp DNA sequences which are centered on the 200-bp bins. In terms of the labels of those sequences, they were generated by collecting profiles from ENCODE and Roadmap Epigenomics data releases, which resulted in a 919 binary vector for each sequence (690 transcription factor binding profiles, 125 DNase I–hypersensitive profiles and 104 histone-mark profiles). To encode the DNA sequence string into a  mathematical form which can be fed to the model, we use the one-hot encoding as discussed in Section \ref{sub:rna_protein_binding_sites_prediction_with_cnn}. In terms of the model, because for DNA sequences, not only do the specific motifs matter, but also the interaction between the upstream and downstream motifs also plays important roles in determining the sequence functionality, we combine CNN, as shown in Fig. \ref{fig:convolution}, and RNN, as shown in Fig. \ref{fig:recurrent}, stacking a bi-directional LSTM layer on top of 1D convolutional layers. The original implementation of \cite{quang2016danq} requires Theano, which has been discontinued. We reimplemented the idea solely using Keras.

This example can be adopted to perform other important predictions, which are similar to this problem, for DNA and RNA sequences, such as DNA methylation state prediction \cite{angermueller2017deepcpg} and long non-coding RNA function prediction \cite{mercer2009long}.

\subsection{Biomedical image classification using transfer learning and ResNet} 
\label{sub:biomedical_image_classification_using_transfer_learning_and_resnet}
The classification of biomedical data has a broad range of applications in biomedical diagnosis. For example, the classification of biomedical images has been used to assist the diagnosis of skin cancer \cite{RN773} and retinal diseases \cite{kermany2018identifying}, which even reaches the expert-level performance. Researchers have also used deep learning to classify the electronic health record (EHR), predicting several medical events, such as, in-hospital mortality, which set the new state-of-the-art performance \cite{rajkomar2018scalable}. However, on the other hand, although deep learning has shown great capability of dealing with the image data, it is highly data-hungry. For example, the famous ImageNet dataset contains over 10 million images \cite{imagenet_cvpr09}. In terms of biomedical data, because of technology limitation and privacy issues, that amount of data is usually not available. In order to maximize the power of deep learning in dealing with biomedical data, transfer learning \cite{yosinski2014transferable} is usually used. Transfer learning is an important machine learning direction, which has its own field \cite{pan2010survey}. But for deep learning models, people usually perform transfer learning in the following way. They first train a standard deep learning model for image classification on the ImageNet dataset. After that, people use the real dataset to fine-tune the model. When fine-tunning the model, they freeze the weights of convolutional layers and use a new fully-connected layer, which has the same number of nodes as the real application's classes, to substitute the last fully-connected layer in the standard model. Then, they train the weights of that new layer. Those frozen convolutional layers are considered as the feature extractor and the newly added fully-connected layer can be considered as the classifier. Although the biomedical images can have some differences from the ImageNet dataset, they usually share similar features. Using transfer learning, the model's performance can be improved significantly \cite{kermany2018identifying}.

In this example, we use the chest X-ray dataset from \cite{kermany2018identifying}. We use Keras to implement this example. We first load the ResNet, whose basic idea is shown in Fig. \ref{fig:architecture} (C), trained on ImageNet, and then freeze all the layer's weights, except for the last 4 layers'. We substitute the last layer with a new layer containing 2 nodes, since that dataset has two classes. Besides, we resize the chest X-ray images to make them the same dimensionality as the original ResNet input image using bilinear interpolation. Finally, we run the standard optimization procedure with cross-entropy loss and Adam optimizer. In order to prevent overfitting, like what we have done in the previous examples, we use dropout \cite{RN586} combined with batch normalization \cite{RN635} and weight decay.

This example can be easily adopted for other biomedical image processing and classification applications \cite{litjens2017survey}, such as breast cancer classification \cite{spanhol2016dataset}, CT image classification \cite{kumar2015lung} and fMRI imaging classification \cite{sarraf2016classification}.

\subsection{Graph embedding for novel protein interaction prediction using GCN} 
\label{sub:graph_embedding_for_novel_protein_interaction_prediction_using_GNN}
In this example, we use graph embedding to handle protein-protein interaction (PPI) networks \cite{han2004evidence}. Specifically, we want to predict novel protein-protein interaction in the yeast PPI network. Understanding protein-protein interaction can be helpful for predicting the functionality of uncharacterized proteins and designing drugs \cite{scott2016small,rual2005towards}. Recently, because the development of large-scale PPI screening techniques, such as the yeast two-hybrid assay \cite{ito2001comprehensive}, the amount of PPI data has increased greatly. Despite the development of those techniques, the current PPI network is still incomplete and noisy, since there are so many proteins existing and so do the complicated interactions. Using computational methods to predict the interaction can be a convenient way to discover novel and important protein-protein interactions \cite{grover2016node2vec}.

We use the yeast protein-protein interaction network as the dataset. In terms of the model, we utilize the graph neural network \cite{kipf2016semi} described in Fig. \ref{fig:gnn}, which can be used to learn the node (protein) embedding based on the network topology. To define the interaction from node embedding, we use the inner product of the embeddings of two nodes as the interaction operation. The higher the product is, the more confidently we believe they have an interaction. Since we have already known some interactions within the network, that is, the existing edges, we can train the graph neural network in a supervised way, using the existing edges (and of course, some protein pairs which do not have interactions as the negative training data) as targets and the cross-entropy as the loss function. After running the optimization until the model converges, we can obtain stable node embeddings for each node based on the network topology.  Then we apply the interaction operation (inner product) to each pair of nodes. If the result is higher than a certain threshold, we will predict that there is an interaction between the two nodes. We implemented this graph example using Tensorflow. 

Notice that the graph data are very common in bioinformatics, such as symptoms-disease networks \cite{zhou2014human}, gene co-expression networks \cite{yang2014gene}, and cell system hierarchy \cite{ma2018using}. This example can be easily adopted for the other network problems.

\subsection{Biology image super-resolution using GAN} 
\label{sub:biology_image_super_resolution_using_gan}
Although, currently, the usage of GAN is still limited to image processing and structure reconstruction \cite{DLBI,chen2018brain}, as a generative model, it has the potential for designing new molecules and drugs \cite{sanchez2018inverse,rampasek2017dr}. To achieve this goal, researchers need to make much more efforts. In this example, we will introduce how to use GAN to perform a legendary but useful task: image super-resolution. In biology, due to the limitation of the imaging acquiring technology, the obtained images are often noisy and of low resolution, such as Cryo-EM images \cite{merk2016breaking} and fluorescence images \cite{RN492}, which requires post-processing to obtain high-resolution, noise-free images. The target of image super-resolution is to output high resolution images given low resolution ones. For example, the input image may only have the resolution as 60 by 60 but the output image can have a resolution as 240 by 240. We also require the super-resolution image to have more true details but less fake ones. This image super-resolution technique has achieved great success in fMRI image fast acquisition \cite{chen2018brain} and fluorescence microscopy super-resolution \cite{DLBI}.

As discussed in Section \ref{sub:generative_model_gan_and_vae} and shown in Fig. \ref{fig:gan}, we train a pair of deep learning models: a generator network and a discriminator network. Given the low-resolution images, the generator network outputs the super-resolution images. The discriminator tries to distinguish the model-generated super-resolution images and the actual high-resolution ones. The discriminator network competes with the generator network so that it can push the generator network to produce the super-resolution images as real as possible. During training, we train both the two networks at the same time to make both of them better. Our ultimate goal is to make the generator network output super-resolution images very close to the real high-resolution ones even if the input low-resolution images are previously unseen in the training data. In terms of the dataset, we use the DIV2K dataset \cite{Agustsson_2017_CVPR_Workshops}. We use ResNet \cite{RN680} as the generator and VGGNet \cite{RN567} as the discriminator. We also add the perceptual loss to the loss function, which stabilizes the high level representation of the super resolution images \cite{johnson2016perceptual}.  We use Tensorflow combined with a higher-level package, Tensorlayer \cite{dong2017tensorlayer}, to implement the GAN idea. 

The example can be adopted for other biological or biomedical image processing applications, such as Cryo-EM images \cite{merk2016breaking} and Cryo-ET images \cite{grunewald2003three}. If data are available, this example probably can also be used for designing new proteins and new drugs \cite{sanchez2018inverse,rampasek2017dr}.

\subsection{High dimensional biological data embedding and generation with VAE} 
\label{sub:high_dimension_biological_data_embedding_and_generation_with_vae}
As GAN, VAE is also a generative model. Thus, theoretically, VAE can also do what GAN can do. As discussed in Section \ref{sub:generative_model_gan_and_vae}, after we approximate the data distribution in the bottleneck layers, we can sample new data from the approximate distribution. Combining the idea of VAE and optimization, we are able to design more efficient drugs \cite{sanchez2018inverse,rampasek2017dr}. At the same time, since the bottleneck layer's dimension is usually much smaller than the original input, this model can also be used to perform dimensionality reduction and mine the important features within the original input \cite{way2017extracting}.

In this example, we briefly show how to use VAE to perform dimensionality reduction for gene expression data \cite{way2017extracting}. We use the dataset from \cite{way2017extracting}, preprocessed from the TCGA database, which collects the gene expression data for over 10,000 different tumors. Within the database, the RNA-seq data describe the high-dimensional state of each tumor. In the dataset we use, the dimensionality is 5,000 for each tumor. Using VAE, we are able to reduce the dimensionality to 100. In that space, it is easier for us to identify the common patterns and signatures between different tumors. We used Keras to implement VAE. For the loss, we use mean-squared error. 

In addition to gene expression data \cite{pierson2015zifa}, this example can be adopted to other applications, which encounter the high dimensionality issue, such as protein fingerprints \cite{das2006low}. In terms of drug design, VAE will have a bright future in this direction, although more efforts need to be made \cite{sanchez2018inverse}.

\section{Perspectives: limitations and suggestions}
\label{sec:discussion}
Currently, there are some difficulties that are frequently encountered when using deep learning. Here we list some of them and give the corresponding suggestions. \textcolor{black}{We also list some related review papers in Table \ref{tab:review}.}

\subsection{Lack of data} 
\label{sub:lacking_of_data}
As we know, deep learning is very data-hungry, since it also contains the representation learning \cite{li2018decision}. To obtain a deep learning model with good performance, we usually need much more data than the shallow algorithms. Besides, the more data we have, the better performance the deep learning model may achieve. Although under most circumstances, the biological data are enough to obtain a good deep learning model \cite{RN561}, sometimes the data may not be enough to directly use deep learning \cite{kermany2018identifying}. There are three suggested ways to deal with this difficulty. Firstly, we can collect the data from similar tasks and use the idea of transfer learning. Although the related data will not increase the amount of real data directly, those data can help learn a better mapping function and a better representation of the original input \cite{yosinski2014transferable}, and thus boost the performance of the model. On the other hand, we can also use a well trained model from another similar task and fine tune the last one or two layers using the limited real data (one example is shown in Section \ref{sub:biomedical_image_classification_using_transfer_learning_and_resnet}). 
\textcolor{black}{A review of the transfer learning methods in the deep learning field can also be referred to \cite{tan2018survey}.}
Secondly, we can perform data augmentation \cite{perez2017effectiveness}. This task is very useful for image data, because rotation, mirroring, and translation of an image usually do not change the label of the image. However, we should be careful when using this technique for bioinformatic data. For example, if we mirror an enzyme sequence, the resulted sequence may no longer be an enzyme sequence. Thirdly, we can consider using simulated data to increase the amount of training data. Sometimes, if the physical process behind the problem is well-known, we can build simulators based on the physical process, which can result in as much simulated data as we want. \cite{DLBI} gives us an example of handling the data requirement for deep learning using simulation.

\subsection{Overfitting} 
\label{sub:overfitting}
Since deep learning models have very high model complexity, with huge amount of parameters which are related in a complex way, the models have high risks of getting overfitted to the training data and being unable to generalize well on the testing data \cite{RN586}. 
\textcolor{black}{Although this problem is not specific to the application of deep learning into bioinformatics, it comes along with almost every deep learning model, which should be fully considered and properly handled when adopting deep learning methods.}
Though recent studies \cite{RN648,RN727,li2018decision} suggest that the implicit bias of the deep learning training process helps model get rid of serious overfitting issues, we still need some techniques to deal with the overfitting issue. Over the years, people have proposed a number of algorithms to alleviate the overfitting issue in deep learning, which can be classified into three types. The first type of techniques, which contains the most famous ones,  acts on the model parameters and the model architecture, including dropout \cite{RN586}, batch normalization \cite{RN635} and weight decay \cite{krogh1992simple}. Dropout is the most well-known technique in the deep learning field. It regularizes the network by randomly discarding nodes and connections during training, which prevents the parameters from co-adapting to the training data too much \cite{RN586}. Although batch normalization \cite{RN635} was first proposed to deal with the internal covariate shift and accelerate the training process, it is also proved to be able to ease the overfitting issue \cite{RN635}. Weight decay \cite{krogh1992simple}, as a universal regularizer, which is widely used in almost all the machine learning algorithms, is also one of the default techniques in the deep learning field. The second category acts on the inputs of the model, such as data augmentation and data corruption \cite{maaten2013learning}. One of the reasons that deep learning models are prone to overfitting is that we do not have enough training data so that the learned distribution may not reflect the actual distribution. Data augmentation increases the amount of training data explicitly and marginalized data corruption \cite{maaten2013learning} can help solve the problem without augmenting the data explicitly. The last type acts on the output of the model. Recently, a method \cite{pereyra2017regularizing} was proposed to regulate the model by penalizing the over-confident outputs, which was shown to be able to regulate both CNN and RNN. \textcolor{black}{A systematic review of regularization methods used to combat the overfitting issue in deep learning can be referred to \cite{kukavcka2017regularization}.}

\subsection{Imbalanced data} 
\label{sub:unbalanced_data}
The biological data are usually imbalanced, with the positive samples being largely outnumbered by the negative ones \cite{yang2011sample}. For example, the number of non-enzyme proteins is much larger than that of a certain type of enzyme proteins \cite{RN775}. The data imbalancing issue also appears in transcription starting site prediction \cite{umarov2017recognition}, Poly(A) site prediction \cite{RN683}, \textit{etc.}. Using the imbalanced data to train a deep learning model may result in undesirable results. For example, the model might predict all the test data with the label from the largest class: all the data, which is composed of 99 negative and 1 positive samples, are predicted to be negative. Although the model's performance is excellent if evaluated using accuracy (as high as $99\%$), the result is in fact terrible as we consider the performance on the small class. To solve the issue, we can use the following techniques. Firstly, we need to use the right criteria to evaluate the prediction result and the loss. For the imbalanced data, we want the model to perform well not only on the large classes but also on the smaller ones. As a result, we should use AUC as the criteria and the corresponding loss \cite{wang2015auc}. Secondly, if we still want to use the cross entropy loss, we can use the weighted cross entropy loss which penalizes the model if it performs terribly on the smaller classes. At the same time, we can upsample smaller classes or downsample larger ones when training the model. Finally, since biological systems often have hierarchical label space, we can build models for each hierarchical level to make the data balanced, as shown in \cite{RN775}.
\textcolor{black}{For the readers' reference, \cite{buda2018systematic} investigates the impact of data imbalance to deep learning model's performance comprehensively, with a comparison of frequently used techniques to alleviate the problem, although those techniques are not specific to biological problems.}

\subsection{Interpretability} 
\label{sub:interpretability}
Although deep learning methods are sometimes criticized for acting like a black-box, it is in fact interpretable \cite{RN654}. In the bioinformatics field, we usually want to interpret deep learning in a way that we can know the useful patterns and motifs detected by the model. 
\textcolor{black}{For example, after building a model for predicting the DNA-protein binding affinity, we may be curious about which motifs on DNA contribute more to the binding affinity landscape \cite{RN730}. Training a deep learning model to perform disease diagnosis, we not only require the prediction and diagnosis results but also want to know how the model makes the decisions and based on which evidences, which can increase our confidence on the model's predictions \cite{choi2016retain}. }
To achieve that, we can assign example-specific importance score for each part of a specific example. In that direction, people can use perturbation-based approaches or backpropagation-based methods \cite{ching2018opportunities}. Perturbation-based approaches \cite{RN612,zhou2015predicting,umarov2017recognition,RN709} change a part of the input and observe its impact on the model's output. Although this idea is easy to understand, it has high computational complexity. Backpropagation-based methods \cite{RN686,shrikumar2017learning} propagate backward the signal from the output layer to the input layer for checking the importance score of different parts of the input. This kind of methods is proved to be useful and under active development \cite{sundararajan2017axiomatic}. Model interpretability can have various meanings in different scenarios. For more comprehensive discussion of deep learning interpretability and all the methods that have been used in bioinformatics, one can refer to \cite{RN654} and Section 5.3 in \cite{ching2018opportunities}.

\subsection{Uncertainty scaling} 
\label{sub:scaling}
When using machine learning methods to perform prediction, we usually do not just want a final predicted label but also want a confidence score for each query from the model, showing how confident the model is sure about the prediction \cite{RN433}. That is an important property, no matter in which application scenarios, because the confidence score can prevent us from believing misleading and unreliable predictions. In biology, it can save us time and resources wasted on validating the misleading prediction results.
\textcolor{black}{
The uncertainty scaling is even more important in healthcare, which can help us evaluate the reliability of machine learning-based disease diagnosis and an automatic clinical decision \cite{leibig2017leveraging}. }
The probability score from the direct deep learning Softmax output is usually not in the right scale, since deep learning models can output overconfident prediction \cite{pereyra2017regularizing}. To obtain reliable probability scores, we need to perform post-scaling to the Softmax output. Several methods have been proposed to output the probability score in the right scale, such as the legendary Platt scaling \cite{RN433}, histogram binning \cite{zadrozny2001obtaining}, isotonic regression \cite{RN439}, Bayssian Binning into Quantiles (BBQ) \cite{naeini2015obtaining}. Recently, temperature scaling was proposed for the deep learning methods specifically, which was shown to be much better than the other methods \cite{RN739}.

\subsection{Catastrophic forgetting} 
\label{sub:catastrophic_forgetting}
A plain deep learning model is usually unable to incorporate new knowledge without interfering the learned knowledge, which is known as catastrophic forgetting \cite{RN737}. For example, after we train a model which can classify 1,000 types of flowers, the 1,001st class of flowers come in. If we only fine-tune the model with the new data, the fine-tuned model's performance on the older classes will be unacceptable \cite{RN798}. This scenario is actually very common in biology since the biological data are alway accumulating and updating. 
\textcolor{black}{For example, the number of entries in PDB \cite{berman2006protein} has increased from 13,590 in 2000 to 147,595 in 2018. The size of Swiss-Prot \cite{bairoch2000swiss} has also increased from around 100,000 in 2000 to 559,077 in 2018.}
With new data being generated, it is very likely that in the future, we will have new classes
\textcolor{black}{as shown in the Enzyme Commission (EC) number system \cite{bairoch2000enzyme}}. 
Although training a completely new model from scratch using both new data and old data can be a straightforward solution, it is computationally intensive and time-consuming to do so, which can also make the learned representation of the original data unstable. Currently, there are three kinds of machine learning methods to deal with the problem based on the neurophysiological theories of human brain \cite{RN724,RN318}, which is free of catastrophic forgetting. The first kind of methods is based on regularizations, such as EWC \cite{RN737}. The second kind of methods uses the dynamic neural network architecture and rehearsal training methods, such as iCaRL \cite{RN798,RN681}. And the last kind of models is based on dual-memory learning systems \cite{dula_weights}. More details can be referred to \cite{RN808,RN798}

\subsection{Reducing computational requirement and model compression} 
\label{sub:reducing_computation_requirement_and_model_compression}
Because deep learning models are usually very complex and have lots of parameters to be trained, it is often computationally demanding and memory intensive to obtain well-trained models and even for the productive usage of the models \cite{cheng2017survey}. Those requirements seriously limit the deployment of deep learning in machines with limited computational power, \textcolor{black}{especially in the field of bioinformatics and healthcare, which is also data intensive. For example, as illustrated in \cite{schatz2010cloud}, the bursting of DNA sequence data even wins over ``Moore's Law'', which requires distributed system or cloud computing to bridge the gap between the computational capability and the tremendous data size. As for the healthcare data, the multiple ways of evaluating people's health and the heterogeneous property of the data have made them more complex, with much larger size \cite{dinov2016volume}, which makes the problem even more computationally demanding \cite{esteva2019guide}. To adopt deep learning methods into those bioinformatics problems which are computational and data intensive, in addition to the development of new hardware devoted to deep learning computing, such as GPUs and FPGAs \cite{zhang2015optimizing}, several methods have been proposed to compress the deep learning model, which can reduce the computational requirement of those models from the beginning. Those methods can be divided into four categories.} 
The first type is parameter pruning, which reduces the redundant parameters that do not contribute to the model's performance significantly, including the famous DeepCompresion \cite{han2015deep}. The second category is knowledge distillation \cite{RN623}, which trains a more compact model with distilled knowledge from the larger model. The third category is to use compact convolutional filters to save parameters \cite{cohen2016group}. And the last category uses low rank factorization \cite{denton2014exploiting} to estimate the informative parameters to preserve. Here we only show the most representative methods for model compression. More comprehensive discussions can be referred to \cite{cheng2017survey}. 


\section{Conclusion}
\label{sec:conclusion}

Deep learning is a highly powerful and useful technique which has facilitated the development of various fields, including bioinformatics. With the advancement of big data era in biology, to further promote the usage of deep learning in bioinformatics, in this review, we first reviewed the achievements of deep learning. After that, we gave a brief and easy-to-understand introduction from shallow neural networks, to legendary convolutional neural networks, legendary recurrent neural networks, graph neural networks, generative adversarial neural networks and variational autoencoder. We also provided detailed examples with implementations to facilitate researchers in adopting and developing their own methods which are based on deep learning. Finally, we pointed out the common difficulties of using deep learning and provided corresponding suggestions. Although this review does not cover all the aspects of deep learning, such as deep reinforcement learning \cite{mnih2015human,RN747,mnih2016asynchronous} and the theoretical aspect of deep learning \cite{RN727,RN648,li2018decision,RN720}, it covers most aspects of the deep learning applications in bioinformatics. We believe this review will shed light on the future development and application of deep learning in bioinformatics, \textcolor{black}{as well as biomedicine \cite{wainberg2018deep} and healthcare \cite{esteva2019guide}}.  

\section*{Acknowledgements}
The research reported in this publication was supported by funding from King Abdullah University of Science and Technology (KAUST), under award number FCC/1/1976-18-01, FCC/1/1976-23-01, FCC/1/1976-25-01, FCC/1/1976-26-01, URF/1/3007-01-01, and URF/1/3450-01-01.

\bibliographystyle{plain}
\bibliography{reference}

\end{document}